\documentstyle[11pt,epsf]{article}
\newtheorem{theorem}{Theorem}

\newcommand {\eqd} {\stackrel{\Delta} {=}}
\newcommand {\exe} {\stackrel{\cdot} {=}}
\newcommand {\lexe} {\stackrel{\cdot} {\leq}}

\newcommand {\bx} {\mbox{\boldmath $x$}}
\newcommand {\by} {\mbox{\boldmath $y$}}

\newcommand {\bE} {\mbox{\boldmath $E$}}

\newcommand {\bX} {\mbox{\boldmath $X$}}

\newcommand{\calA}{{\cal A}}

\newcommand{\calC}{{\cal C}}

\newcommand{\calE}{{\cal E}}

\newcommand{\calG}{{\cal G}}

\newcommand{\calR}{{\cal R}}

\newcommand{\calX}{{\cal X}}
\newcommand{\calY}{{\cal Y}}

\begin{document}
\thispagestyle{empty}
\title{Error Exponents of Erasure/List Decoding Revisited
via Moments of Distance Enumerators
%\thanks{This research was supported by my wife and kids.}
}
\author{Neri Merhav
%\thanks{
%Currently on sabbatical leave at HP Laboratories,
%1501 Page Mill Road, MS 3U-4, Palo Alto CA 94304, USA.}
}
%\date{}
\maketitle

\begin{center}
Department of Electrical Engineering \\
Technion - Israel Institute of Technology \\
Haifa 32000, ISRAEL \\
\end{center}
\vspace{1.5\baselineskip}
\setlength{\baselineskip}{2\baselineskip}

\begin{abstract}
The analysis of random coding
error exponents pertaining to erasure/list decoding, due to Forney, is revisited.
Instead of using Jensen's 
inequality as well as some other inequalities in the derivation, we demonstrate that an
exponentially tight analysis can be carried out by assessing the relevant moments of a certain
distance enumerator. The resulting bound has the following advantages: (i) it is at least as
tight as Forney's bound, (ii) under certain symmetry conditions associated with the channel and
the random coding distribution, it is simpler than Forney's bound in the sense that it involves
an optimization over one parameter only (rather than two), and (iii) in certain special cases,
like the binary symmetric channel (BSC), the optimum value of this parameter can be found in
closed form, and so, there is no need to conduct a numerical search. We have not found yet, however,
a numerical example where this new bound is strictly better than Forney's bound. This may provide an
additional evidence to support Forney's conjecture that his bound is tight for the average code.
We believe that the technique we suggest in this paper can be useful in simplifying, and hopefully
also improving, exponential error bounds in other problem settings as well.

\vspace{0.25cm}

\noindent
{\bf Index Terms:} random coding, erasure, list, error exponent, distance enumerator.
\end{abstract}

\section{Introduction}

In his celebrated paper \cite{Forney68}, Forney extended Gallager's bounding techniques
\cite{Gallager68} and found exponential error bounds for the ensemble performance of
optimum generalized decoding rules that include the options of erasure, variable size lists, and
decision feedback (see also later studies, e.g., 
\cite{ACZ96}, 
\cite{Hashimoto99},\cite{HT97},
\cite{KNG07},
\cite{MF07}, and
\cite{Viterbi69}). 

Stated informally, Forney \cite{Forney68} considered a communication system where
a code of block length $n$ and size $M=e^{nR}$ ($R$ being the coding rate), 
drawn independently at random under
a distribution $\{P(x)\}$, is used for a discrete memoryless channel (DMC) 
$\{P(y|x)\}$ and decoded
with an erasure/list option. For the erasure case, in which we focus hereafter, an optimum
tradeoff was sought between the probability of erasure (no decoding) and the probability
of undetected decoding error. This tradeoff is 
optimally controlled by a threshold parameter $T$ 
of the function $e^{nT}$ to which
one compares the ratio between the likelihood of each 
hypothesized message and the sum of likelihoods of all other messages.
If this ratio exceeds $e^{nT}$ for some message, 
a decision is made in favor of that message,
otherwise, an erasure is declared.

Forney's main result in \cite{Forney68}
is a single--letter lower bound $E_1(R,T)$ to the exponent of the probability
of the event $\calE_1$ of not making the correct decision, namely, either
erasing or making the wrong decision. This lower bound is given by
\begin{equation}
\label{forneyerrexp}
E_1(R,T)=\max_{0\le s\le \rho\le 1}[E_0(s,\rho)-\rho R-sT]
\end{equation}
where
\begin{equation}
E_0(s,\rho)=-\ln\left[\sum_y\left(\sum_xP(x)P^{1-s}(y|x)\right)\cdot\right.
\left.\left(\sum_{x'}P(x')P^{s/\rho}(y|x')\right)^\rho\right].
\end{equation}
The probability of the undetected error event $\calE_2$ (i.e., the event of not erasing but making
a wrong estimate of the transmitted message) is given by $E_2(R,T)=E_1(R,T)+T$.\footnote{
Forney also provides improved (expurgated) exponents at low rates, but we will focus here 
solely on (\ref{forneyerrexp}).}
As can be seen, the computation of $E_1(R,T)$ involves an optimization
over two auxiliary parameters, $\rho$ and $s$, which in general requires a 
two--dimensional search over these two parameters by some method. This is
different from Gallager's random coding error exponent function for ordinary decoding
(without erasures), which is given by:
\begin{equation}
\label{gallagererrexp}
E_r(R)=\max_{0\le \rho\le 1}[E_0(\rho)-\rho R],
\end{equation}
with $E_0(\rho)$ being defined as
\begin{equation}
E_0(\rho)=-\ln\left[\sum_y\left(\sum_xP(x)P^{1/(1+\rho)}(y|x)\right)^{1+\rho}
\right],
\end{equation}
where there is only one parameter to be optimized.
In \cite{Forney68}, one of the steps in the derivation involves
the inequality $(\sum_i a_i)^r\le 
\sum_i a_i^r$, which holds 
for $r\le 1$ and non--negative $\{a_i\}$ (cf.\ eq.\ (90) in \cite{Forney68}),
and another step (eq.\ (91e) therein) applies Jensen's inequality. The former inequality
introduces an additional parameter, denoted $\rho$, to be optimized 
together with the original parameter, $s$.\footnote{The parameter $s$
is introduced, as in many other derivations, as the
power of the likelihood ratio that bounds the indicator function of the error event.
This point will be elaborated on in Section 4.}

In this paper, we offer a different technique for deriving a lower bound to the
exponent of the probability of $\calE_1$, which avoids 
the use of these inequalities. Instead,
an exponentially tight evaluation of the relevant expression
is derived by assessing the moments of a certain distance enumerator, 
and so, the resulting bound is at least as tight as Forney's bound.
Since the first above--mentioned inequality
is bypassed, there is no need for the additional parameter $\rho$, and so, under certain
symmetry conditions (that often hold) on the
random coding distribution and the channel, the resulting bound 
is not only at least as tight as Forney's bound, but it is also
simpler in the sense that there is only
one parameter to optimize rather than two. Moreover, this optimization can be carried
out in closed form at least in some special cases like the binary symmetric channel (BSC).
We have not found yet, however,
a convincing\footnote{In a few cases, small differences were found, but these could
attributed to insufficient resolution of the two--dimensional search for the optimum
$\rho$ and $s$ in Forney's bound.}
numerical example where the new bound is {\it strictly} 
better than Forney's bound. This may serve as an
additional evidence to support Forney's conjecture that his bound is tight for the average code.
Nevertheless, the question whether there exist situations where the new bound is strictly
better, remains open.

We wish to emphasize that
the main message of this contribution, is not merely
in the simplification of the error exponent bound in this specific problem of
decoding with erasures,
but more importantly, in the analysis technique we offer here, which, we believe, is applicable
to quite many other problem settings as well. It is conceivable that in some of these problems,
the proposed technique could not only simply, but perhaps also improve on currently known bounds.
The underlying ideas behind this technique are inspired from
the statistical mechanical point of view on random code ensembles,
offered in \cite{MM06} and further elaborated on
in \cite{p117}.

The outline of this paper is as follows. In Section 2,
we establish notation conventions and give some necessary
background in more detail. In Section 3, we present the main result
along with a short discussion. Finally, in Section 4, 
we provide the detailed derivation of the new bound, first for the special
case of the BSC, and then more generally.

\section{Notation and Preliminaries}

Throughout this paper, scalar random variables (RV's) will be denoted by capital
letters, their sample values will be denoted by
the respective lower case letters, and their alphabets will be denoted
by the respective calligraphic letters.
A similar convention will apply to
random vectors of dimension $n$ and their sample values,
which will be denoted with same symbols in the bold face font.
The set of all $n$--vectors with components taking values in a certain finite alphabet,
will be denoted as the same alphabet superscripted by $n$.
Thus, for example, a random vector $\bX=(X_1,\ldots,X_n)$ may assume
a specific vector value $\bx=(x_1,\ldots,x_n)\in\calX^n$ as each
component takes values in $\calX$.
Sources and channels will be denoted generically by the letter $P$ or $Q$.
Information theoretic quantities entropies and conditional entropies,
will be denoted following the usual conventions
of the information theory literature, e.g., $H(X)$, $H(X|Y)$, and so on.
When we wish to emphasize the dependence of the
entropy on a certain underlying probability distribution, say $Q$, we subscript it by 
$Q$, i.e., use notations like $H_Q(X)$, $H_Q(X|Y)$, etc.
The expectation operator will be denoted by $\bE\{\cdot\}$, and once again,
when we wish to make the dependence on the underlying distribution $Q$ clear, we denote it
by $\bE_Q\{\cdot\}$.
The cardinality of a finite set $\calA$ will be denoted by $|\calA|$.
The indicator function of an event $\calE$ will be denoted by $1\{\calE\}$.
For a given sequence $\by\in\calY^n$, $\calY$ being a finite alphabet, $\hat{P}_{\by}$
will denote the empirical distribution on $\calY$ extracted from $\by$, in other words,
$\hat{P}_{\by}$ is the vector $\{\hat{P}_{\by}(y),~y\in\calY\}$, where $\hat{P}_{\by}(y)$
is the relative frequency of the letter $y$ in the vector $\by$.
For two sequences of positive numbers, $\{a_n\}$ and $\{b_n\}$, the notation
$a_n\exe b_n$ means that $\{a_n\}$ and $\{b_n\}$
are of the same exponential order, i.e., $\frac{1}{n}\ln\frac{a_n}{b_n}\to 0$
as $n\to\infty$. Similarly, $a_n\lexe b_n$ means that 
$\limsup_{n\to\infty}\frac{1}{n}\ln\frac{a_n}{b_n}\le 0$, and so on.

Consider a discrete memoryless channel (DMC) with a finite
input alphabet $\calX$,
finite output alphabet $\calY$, and single--letter transition probabilities
$\{P(y|x),~x\in\calX,~y\in\calY\}$. As the channel is fed by an input vector
$\bx\in\calX^n$, it generates an output vector
$\by\in\calY^n$ according to the sequence conditional probability distributions
\begin{equation}
P(y_i|x_1,\ldots,x_i,y_1,\ldots,y_{i-1})=P(y_i|x_i), ~~~i=1,2,\ldots,n
\end{equation}
where for $i=1$, $(y_1,\ldots,y_{i-1})$ is understood as the null string.
A rate--$R$ block code of length $n$ consists of $M=e^{nR}$ $n$--vectors
$\bx_m\in\calX^n$, $m=1,2,\ldots,M$, which represent $M$ different messages.
We will assume that all possible messages are
a--priori equiprobable, i.e., $P(m)=1/M$ for all $m=1,2,\ldots,M$.

A decoder with an erasure option is a partition of $\calY^n$ into $(M+1)$ regions,
$\calR_0,\calR_1,\ldots,\calR_M$. Such a decoder works as follows: If $\by$ falls
into $\calR_m$, $m=1,2,\ldots,M$, then a decision is made in favor of message number $m$.
If $\by\in\calR_0$, no decision is made and an erasure is declared. We will refer to
$\calR_0$ as the {\it erasure event}.
Given a code $\calC=\{\bx_1,\ldots,\bx_M\}$
and a decoder $\calR=(\calR_0,\calR_1,\ldots,\calR_m)$, let us now define
two additional undesired events.
The event $\calE_1$ is the event of not making the right decision. This
event is the disjoint union of the erasure event and the event
$\calE_2$, which is the
{\it undetected error} event, namely, the event of making the wrong decision.
The probabilities of all three events are defined as follows:
\begin{eqnarray}
\mbox{Pr}\{\calE_1\}&=&
\sum_{m=1}^M\sum_{\by\in\calR_m^c}P(\bx_m,\by)=
\frac{1}{M}\sum_{m=1}^M\sum_{\by\in\calR_m^c}P(\by|\bx_m)\\
\mbox{Pr}\{\calE_2\}&=&
\sum_{m=1}^M\sum_{\by\in\calR_m}\sum_{m'\ne m}P(\bx_{m'},\by)=
\frac{1}{M}\sum_{m=1}^M\sum_{\by\in\calR_m}\sum_{m'\ne m}P(\by|\bx_{m'})\\
\mbox{Pr}\{\calR_0\}&=&\mbox{Pr}\{\calE_1\}-\mbox{Pr}\{\calE_2\}.
\end{eqnarray}
Forney \cite{Forney68}
shows, using the Neyman--Pearson theorem,
that the best
tradeoff between $\mbox{Pr}\{\calE_1\}$ and $\mbox{Pr}\{\calE_2\}$ (or, equivalently,
between $\mbox{Pr}\{\calR_0\}$ and $\mbox{Pr}\{\calE_2\}$) is attained by the decoder
$\calR^*=(\calR_0^*,\calR_1^*,\ldots,\calR_M^*)$ defined by
\begin{eqnarray}
\calR_m^*&=&\left\{\by:~\frac{P(\by|\bx_m)}{\sum_{m'\ne m}
P(\by|\bx_{m'})}\ge e^{nT}\right\},~~m=1,2,\ldots,M\nonumber\\
\calR_0^*&=&\bigcap_{m=1}^M (\calR_m^*)^c,
\end{eqnarray}
where $(\calR_m^*)^c$ is the complement of $\calR_m^*$, and
where $T\ge 0$ is a parameter, henceforth referred to as the {\it threshold},
which controls the balance between the probabilities of $\calE_1$ and $\calE_2$.
Forney devotes the remaining part of his paper \cite{Forney68}
to derive lower bounds, as well as to investigate properties,
of the random coding
exponents (associated with $\calR^*$),
$E_1(R,T)$ and $E_2(R,T)$, of $\overline{\mbox{Pr}}\{\calE_1\}$ 
and $\overline{\mbox{Pr}}\{\calE_2\}$,
the average
probabilities of $\calE_1$ and $\calE_2$, respectively,
(w.r.t.) the ensemble of randomly selected codes,
drawn independently according to an i.i.d.\ distribution $P(\bx)=\prod_{i=1}^nP(x_i)$.
As mentioned in the Introduction, $E_1(R,T)$ is given by (\ref{forneyerrexp}) and
$E_2(R,T)=E_1(R,T)+T$.

\section{Main Result}

Our main result in this paper is the following:
\begin{theorem}
Assume that the random coding distribution $\{P(x),~x\in\calX\}$
and the channel transition matrix $\{P(y|x),~x\in\calX,~y\in\calY\}$
are such that for every real $s$,
\begin{equation}
\label{condition}
\gamma_y(s)\eqd-\ln \left[\sum_{x\in\calX} P(x)P^s(y|x)\right]
\end{equation}
is independent of $y$, in which case, it will be denoted by $\gamma(s)$.
Let $s_R$ be the solution to the equation
\begin{equation}
\gamma(s)-s\gamma'(s)=R,
\end{equation}
where $\gamma'(s)$ is the derivative of $\gamma(s)$.
Finally, let 
\begin{equation}
E_1^*(R,T,s)=\Lambda(R,s)+\gamma(1-s)-sT-\ln|\calY|
\end{equation}
where
\begin{equation}
\Lambda(R,s)=\left\{\begin{array}{ll}
\gamma(s)-R & s \ge s_R\\
s\gamma'(s_R) & s < s_R \end{array}\right.
\end{equation}
Then,
\begin{equation}
\overline{\mbox{Pr}}\{\calE_1\}\lexe e^{-nE_1^*(R,T)}
\end{equation}
where $E_1^*(R,T)=\sup_{s\ge 0}E_1^*(R,T,s)$
and
\begin{equation}
\overline{\mbox{Pr}}\{\calE_2\}\lexe e^{-nE_2^*(R,T)}
\end{equation}
where $E_2^*(R,T)=E_1^*(R,T)+T$. Also, $E_1^*(R,T)\ge E_1(R,T)$, where
$E_1(R,T)$ is given in (\ref{forneyerrexp}).
\end{theorem}
Three comments are in order regarding the condition that $\gamma_y(s)$ of
eq.\ (\ref{condition}) is independent of $y$. 

The first comment is that this condition is obviously
satisfied when $\{P(x)\}$ is uniform and the columns of the matrix
$\{a_{xy}\}=\{P(y|x)\}$ are permutations of each other, because then
the summations $\sum_x P(x)P^\beta(y|x)$, for the various $y$'s,
consist of exactly the same terms, just in a different order. This is the 
case, for example, when $\calX=\calY$ is a group endowed with an addition/subtraction
operation (e.g., addition/subtraction modulo the alphabet size), and the channel is additive
in the sense that the `noise' $(Y-X)$ is 
statistically independent of $X$. Somewhat more generally, the condition
$\gamma_y(s)=\gamma(s)$ for all $y$ holds when the different columns of the matrix
$\{P(y|x)\}$ are formed by permutations of each other subject to the following rule:
$P(y|x)$ can be
permuted with $P(y|x')$ if $P(x)=P(x')$.
For example, let $\calX=\{A,B,C\}$ and $\calY=\{1,2\}$,
let the random coding distribution be given by $P(A)=a$, $P(B)=P(C)=(1-a)/2$,
and let the channel be given by $P(0|A)=P(1|A)=1/2$, $P(0|B)=P(1|C)=1-P(0|C)=1-P(1|B)=b$.
In this case,
the condition is satisfied and $\gamma(s)=-\ln[(1-a)/2)(b^s+(1-b)^s)+a\cdot 2^{-s}]$.

The second comment is that
the derivation of the bound, using the proposed technique,
can be carried out, in principle, even without this condition
on $\gamma_y(s)$. In the absence of this condition, one ends up with an exponential expression
that depends, for each $\by$, on the empirical distribution $\hat{P}_{\by}$, and its
summation over $\by$ can then be handled using the method of types, which involves optimization
over $\{\hat{P}_{\by}\}$, or in the limit of large $n$, optimization over the continuum of
probability distributions on $\calY$. But then we are
loosing the simplicity of the bound relative to Forney's bound, since this optimization
is at least as complicated as the optimization over the additional parameter $\rho$ in
\cite{Forney68}.

Our last comment, in the context of this condition on $\gamma_y(s)$, is that even
when it holds, it is not apparent that the expression of Forney's bound $E_1(R,T)$ can be
simplified directly in a trivial manner, nor can we see how the optimum parameters $\rho$
and $s$ can be found analytically in closed form.
This is true even in the simplest case of the BSC.

\section{Derivation of the New Bound}

\subsection{Background}

The first few steps of the derivation
are similar to those in \cite{Forney68}: For a given code
and for every $s\ge 0$,
\begin{eqnarray}
\mbox{Pr}\{\calE_1\}&=&\frac{1}{M}\sum_{m=1}^M\sum_{\by\in (\calR_m^*)^c}P(\by|\bx_m)\nonumber\\
&=&\frac{1}{M}\sum_{m=1}^M\sum_{\by\in\calY^n}P(\by|\bx_m)\cdot 1\left\{\frac{e^{nT}
\sum_{m'\ne m}P(\by|\bx_{m'})}
{P(\by|\bx_m)} \geq 1\right\}\nonumber\\
&\le&\frac{1}{M}\sum_{m=1}^M\sum_{\by\in\calY^n}P(\by|\bx_m)\left(\frac{e^{nT}
\sum_{m'\ne m}P(\by|\bx_{m'})}
{P(\by|\bx_m)}\right)^s\nonumber\\
&=&\frac{e^{nsT}}{M}\sum_{m=1}^M\sum_{\by\in\calY^n}P^{1-s}(\by|\bx_m)
\left(\sum_{m'\ne m}P(\by|\bx_{m'})\right)^s.
\end{eqnarray}
As for $\calE_2$, we have similarly,
\begin{equation}
\mbox{Pr}\{\calE_2\}\le e^{-n(1-s)T}\sum_{\by\in\calY^n}P^{1-s}(\by|\bX_m)
\left(\sum_{m'\ne m}
P(\by|\bX_{m'})\right)^s.
\end{equation}
Since this differs from the bound on $\mbox{Pr}\{\calE_1\}$ only by the constant factor 
$e^{-nT}$, it will be sufficient to focus on $\calE_1$ only.
Taking now the expectation w.r.t.\ the ensemble 
of codes, and using the fact that $\bX_m$ is independent of all other codewords,
we get:
\begin{equation}
\label{startingpoint}
\overline{\mbox{Pr}}\{\calE_1\}\le e^{nsT}\sum_{\by\in\calY^n}\bE\{P^{1-s}(\by|\bX_m)\}\cdot
\bE\left\{\left(\sum_{m'\ne m}
P(\by|\bX_{m'})\right)^s\right\}.
\end{equation}
The first factor of the summand is easy to handle:
\begin{eqnarray}
\bE\{P^{1-s}(\by|\bX_m)\}&=&\sum_{\bx\in\calX^n}P(\bx)P^{1-s}(\by|\bx)\nonumber\\
&=&\prod_{i=1}^n[\sum_{x\in\calX}P(x)P^{1-s}(y_i|x)]\nonumber\\
&=&e^{-n\gamma(1-s)}.
\end{eqnarray}
Concerning the second factor of the summand,
Forney's approach is to use the inequality $(\sum_i a_i)^r\le \sum_i a_i^r$,
which holds when $\{a_i\}$ are positive and $r\le 1$, in order to upper bound
$$\bE\left\{\left(\sum_{m'\ne m}
P(\by|\bX_{m'})\right)^s\right\}$$
by
$$\bE\left\{\left(\sum_{m'\ne m}
P(\by|\bX_{m'})^{s/\rho}\right)^\rho\right\},~~~~\rho \ge s,$$
and then (similarly to Gallager) use Jensen's inequality
to insert the expectation into the bracketed expression,
which is allowed by limiting $\rho$ to be less than unity. In other words,
the above expression is further upper bounded in \cite{Forney68} by
$$\left(\sum_{m'\ne m}\bE\left\{
P(\by|\bX_{m'})^{s/\rho}\right\}\right)^\rho,~~~~\rho \le 1.$$
The idea here, as we understand it, is that the 
parameter $\rho$ controls the tradeoff between the gap
pertaining to the first inequality and the one associated with the second inequality.
The first gap is maximum when $\rho=1$ and non--existent when $\rho=s$ ($s\le 1$), whereas
for the second gap it is vice versa.

We, on the other hand, will use a different route,
where all steps of the derivation will be clearly exponentially tight, and without
introducing the additional parameter $\rho$. 
To simplify the exposition and make it easier to gain some geometrical insight, it will
be instructive to begin with the special case of the BSC 
and the uniform random coding
distribution. The extension to more general DMC's
and random coding distributions will be given in Subsection \ref{general}. 
Readers who are interested in the more general case only may skip to Subsection \ref{general}
without loss of continuity. 

\subsection{The BSC with the uniform random coding distribution}

Consider the special case where $\calX=\calY=\{0,1\}$, the channel is a BSC with
a crossover probability $p$, and the random coding 
distribution is uniform over the Hamming space $\{0,1\}^n$, namely,
$P(\bx)=2^{-n}$ for all $\bx\in\{0,1\}^n$. 
First, concerning the first factor in the summand of (\ref{startingpoint}), we have,
in this special case:
\begin{equation}
\gamma(1-s)=-\ln\left[\frac{1}{2}p^{1-s}+\frac{1}{2}(1-p)^{1-s}\right]=\ln 2-\ln[p^{1-s}+(1-p)^{1-s}].
\end{equation}
As for the second factor, we proceed as follows. Define $\beta=\ln\frac{1-p}{p}$ and for
a given $\by$, let $N_{\by}(d)$ denote distance enumerator relative to $\by$, that is,
the number of incorrect codewords $\{\bx_{m'},~m'\ne m\}$
at Hamming distance $d$ from $\by$. We then have:
\begin{eqnarray}
\bE\left\{\left(\sum_{m'\ne m}
P(\by|\bX_{m'})\right)^s\right\}&=&\bE\left\{\left[(1-p)^n
\sum_{d=0}^nN_{\by}(d)e^{-\beta d}\right]^s\right\}\nonumber\\
&\exe&(1-p)^{ns}\sum_{d=0}^n\bE\{N_{\by}^s(d)\}e^{-\beta sd}.
\end{eqnarray}
The second (exponential) equality is the {\it first main point} in our approach:
It holds, even before taking the expectations,
because the summation over $d$ consists of a {\it subexponential} number of terms, and so, both
$[\sum_d N_{\by}(d)e^{-\beta d}]^s$ and
$\sum_d N_{\by}^s(d)e^{-\beta sd}$ are of the same exponential order as 
$\max_d N_{\by}^s(d)e^{-\beta sd}=
[\max_d N_{\by}(d)e^{-\beta d}]^s$. This is different from the original summation over $m'$,
which contains an {\it exponential} number of terms.
Thus, the key issue here is how to assess the power--$s$ moments 
of the distance enumerator $N_{\by}(d)$.
To this end, we have to distinguish between two ranges of $d$, or equivalently,
$\delta=d/n$.
Let $\delta_{GV}(R)$ denote the normalized Gilbert--Varshamov
(GV) distance, $\delta_{GV}=d_{GV}/n$, i.e., the smaller solution, $\delta$, to the equation 
$$h(\delta)=\ln 2-R,$$
where 
$$h(\delta)=-\delta\ln\delta-(1-\delta)\ln(1-\delta),~~~\delta\in[0,1].$$ 

Now, the {\it second main point} of the proposed analysis approach
is that $\bE\{N_{\by}^s(d)\}$ behaves differently\footnote{The intuition
behind this different behavior is that when $h(\delta)+R-\ln 2 > 0$, the RV $N_{\by}(d)$,
which is the sum of $e^{nR}-1$ many i.i.d.\ binary RV's, $1\{d(\bX_{m'},\by)=d\}$,
concentrates extremely (double--exponentially) 
rapidly around its expectation $e^{n[R+h(\delta)-\ln 2]}$, whereas for $h(\delta)+R-\ln 2 < 0$,
$N_{\by}(d)$ is typically zero, and so, the dominant term of
$\bE\{N_{\by}^s(d)\}$ is $1^s\cdot\mbox{Pr}\{N_{\by}(d)=1\}\approx e^{n[R+h(\delta)-\ln 2]}$.}
for the case $\delta_{GV}(R) \le \delta \le 1-\delta_{GV}(R)$ and for the case
$\delta < \delta_{GV}(R)$ or $\delta > 1-\delta_{GV}(R)$. 
Let us define then $\calG_R=\{\delta:~\delta_{GV}(R)\le \delta \le 1-\delta_{GV}(R)\}$.
In particular, using
the large deviations behavior of $N_{\by}(n\delta)$, $\delta\in[0,1]$, 
as the sum of $e^{nR}-1$ binary i.i.d.\ 
RV's, it is easily seen (see Appendix) that
\begin{equation}
\label{moments}
\bE\{N_{\by}^s(n\delta)\}\exe \left\{\begin{array}{ll}
e^{ns[R+h(\delta)-\ln 2]} & \delta \in \calG_R\\
e^{n[R+h(\delta)-\ln 2]} & \delta \in \calG_R^c .
\end{array}\right.
\end{equation}
Thus,
\begin{eqnarray}
\label{summation}
&&\bE\left\{\left(\sum_{m'\ne m}
P(\by|\bX_{m'})\right)^s\right\}\nonumber\\
&\exe&(1-p)^{ns}
\left[\sum_{\delta\in \calG_R}e^{ns[R+h(\delta)-\ln 2]}\cdot 
e^{-\beta sn\delta}+
\sum_{\delta\in \calG_R^c}e^{n[R+h(\delta)-\ln 2]}
\cdot e^{-\beta sn\delta}\right]\nonumber\\
&\exe&(1-p)^{ns}\left[e^{ns(R-\ln 2)}\cdot
\exp\{ns\max_{\delta\in\calG_R}[h(\delta)-\beta\delta]\}+
e^{n(R-\ln 2)}\cdot\exp\{n\max_{\delta\in\calG_R^c}[h(\delta)-\beta s\delta]\}\right]
\end{eqnarray}
We are assuming, of course,
$R < C= \ln 2-h(p)$, which is equivalent to $p < \delta_{GV}(R)$ or $p > 1-\delta_{GV}(R)$.
We also assume, for the sake of simplicity and without essential
loss of generality, that $p < 1/2$, which will leave us only with
the first possibility of $p < \delta_{GV}(R)$. Therefore, the global 
(unconstrained) maximum of $h(\delta)-\beta\delta$, which is attained at $\delta=p$,
falls outside $\calG_R$, and so, $\max_{\delta\in\calG_R}[h(\delta)-\beta\delta]$
is attained at $\delta=\delta_{GV}(R)$ which yields
$$\max_{\delta\in\calG_R}[h(\delta)-\beta\delta]=h(\delta_{GV}(R))-\beta\delta_{GV}(R)
=\ln 2-R-\beta\delta_{GV}(R).$$
Thus, the first term in the 
large square brackets of the r.h.s.\ of (\ref{summation}) is of the
exponential order of $e^{-ns\beta\delta_{GV}(R)}$.
As for the second term, the unconstrained maximum of $h(\delta)-\beta s\delta$ is obtained
at $\delta=p_s\eqd \frac{p^s}{p^s+(1-p)^s}$, which can be either larger or smaller
than $\delta_{GV}(R)$, depending on $s$. Specifically,
\begin{equation}
\max_{\delta\in\calG_R^c}[h(\delta)-\beta s\delta]=\left\{\begin{array}{ll}
h(p_s)-\beta sp_s & p_s \le \delta_{GV}(R) \\
\ln 2-R-\beta s\delta_{GV}(R) & p_s > \delta_{GV}(R) 
\end{array}\right.
\end{equation}
The condition $p_s\le \delta_{GV}(R)$ is equivalent to
$$s \ge s_R\eqd \frac{\ln[(1-\delta_{GV}(R))/\delta_{GV}(R)]}{\beta}.$$
Thus, the second term in the square brackets of the r.h.s.\ of eq.\ (\ref{summation})
is of the order of $e^{-n\mu(s,R)}$, where
\begin{equation}
\mu(s,R)=\left\{\begin{array}{ll}
\mu_0(s,R) & s \ge s_R\\
\beta s\delta_{GV}(R) & s < s_R
\end{array}\right.
\end{equation}
and where
\begin{eqnarray}
\mu_0(s,R)&=&\beta sp_s-h(p_s)+\ln 2-R\nonumber\\
&=&s\ln(1-p)-\ln[p^s+(1-p)^s]+\ln 2-R.
\end{eqnarray}
Next, observe that the second term, $e^{-n\mu(s,R)}$, is always the dominant term:
For $s < s_R$, 
this is trivial as both terms behave like $e^{-n\beta s \delta_{GV}(R)}$.
For $s \ge s_R$ (namely, $p_s \le \delta_{GV}(R)$), as $\delta=p_s$ 
achieves the {\it global} minimum of the function $f(\delta)\eqd\beta s\delta-h(\delta)+\ln 2-R$,
we have 
$$\mu_0(s,R)=f(p_s)\le f(\delta_{GV}(R))=\beta s\delta_{GV}(R).$$
Therefore, we have established that
\begin{equation}
\bE\left\{\left(\sum_{m'\ne m}
P(\by|\bX_{m'})\right)^s\right\}\exe \exp\left\{-n\left[s\ln\frac{1}{1-p}+
\mu(s,R)\right]\right\}
\end{equation}
independently of $\by$.
Finally, we get:
\begin{equation}
\overline{\mbox{Pr}}\{\calE_1\}\lexe 
e^{nsT}\cdot 2^n\cdot e^{-n[\ln 2-\ln(p^{1-s}+(1-p)^{1-s})]}\cdot 
\exp\left\{-n\left[s\ln\frac{1}{1-p}+
\mu(s,R)\right]\right\}
=e^{-nE_1(R,T,s)}
\end{equation}
where
$$E_1(R,T,s)\eqd \mu(s,R)+s\ln\frac{1}{1-p}
-\ln[p^{1-s}+(1-p)^{1-s}]-sT.$$
It is immediate to verify that this coincides with the
expression of Theorem 1 when specialized to the case of the BSC
with a uniform random coding distribution.

We next derive closed form expressions for the
optimum value of $s$, denoted $s_{\mbox{opt}}$,
using the following consideration:
We have seen that $E_1^*(R,T,s)$ is given by
$$F(s)\eqd\mu_0(s,R)+s\ln\frac{1}{1-p}-\ln[p^{1-s}+(1-p)^{1-s}]-sT$$
for $s\ge s_R$, and by
$$G(s)\eqd\beta s\delta_{GV}(R)+s\ln\frac{1}{1-p}-\ln[p^{1-s}+(1-p)^{1-s}]-sT$$
for $s<s_R$. Both $F(s)$ and $G(s)$ are easily seen to be concave functions
and hence have a unique maximum each, which can be found by equating the
corresponding derivative to zero.
We have also seen that $F(s)\leq G(s)$ for all $s$, with equality
at $s=s_R$ and only at that point. This means that $F(s)$ and $G(s)$ are tangential
to each other at $s=s_R$, in other words, $F(s_R)=G(s_R)$ and $F'(s_R)=G'(s_R)$, where
$F'$ and $G'$ are the derivatives of $F$ and $G$, respectively. Now, there are three
possibilities: If $F'(s_R)=G'(s_R)=0$, then $s_{\mbox{opt}}=s_R$. If $F'(s_R)=G'(s_R) < 0$,
then $s_{\mbox{opt}} < s_R$ is found by solving the equation $G'(s)=0$. If $F'(s_R)=G'(s_R) > 0$,
then $s_{\mbox{opt}} > s_R$ is found by solving the equation $F'(s)=0$. 

Let us assume first that $s_{\mbox{opt}} < s_R$.
Then, the equation $G'(s)=0$ is equivalent to:
$$\beta \delta_{GV}(R)+\ln\frac{1}{1-p}+p_{1-s}\ln p+(1-p_{1-s})\ln(1-p)-T=0$$
or 
$$\beta p_{1-s}=\beta\delta_{GV}(R)-T$$
whose solution is:
\begin{equation}
\label{sstar}
s^*=1-\frac{1}{\beta}\ln\frac{\beta(1-\delta_{GV}(R))+T}{\beta\delta_{GV}(R)-T}.
\end{equation}
Of course, if the r.h.s.\ of (\ref{sstar}) turns out to be negative, then 
$s_{\mbox{opt}}=0$. Thus, overall
\begin{equation}
\label{sstar1}
s_{\mbox{opt}}=s_1(p,R,T)\eqd
\left[1-\frac{1}{\beta}\ln\frac{\beta(1-\delta_{GV}(R))+T}{\beta\delta_{GV}(R)-T}\right]_+,
\end{equation}
where $[x]_+\eqd\max\{x,0\}$. Of course, when $s_{\mbox{opt}}=0$, the new bound $E_1^*(R,T)$
vanishes. 

Next, assume that $s_{\mbox{opt}}>s_R$.
In this case,
\begin{eqnarray}
E_1(R,T,s)&=&F(s)\nonumber\\
&=&\ln 2-\ln[p^s+(1-p)^s]-\ln[p^{1-s}+(1-p)^{1-s}]-R-sT.
\end{eqnarray}
Thus, the optimum $s$ minimizes the convex function
\begin{eqnarray}
f(s)&=&\ln[p^s+(1-p)^s]+\ln[p^{1-s}+(1-p)^{1-s}]+sT\nonumber\\
&=&\ln\left[1+(1-p)\left(\frac{p}{1-p}\right)^s+p\left(\frac{1-p}{p}\right)^s\right]+sT.
\end{eqnarray}
Equating the derivative to zero, we get:
\begin{equation}
f'(s)\equiv\frac{-\left(\frac{p}{1-p}\right)^s\cdot(1-p)\beta
+\left(\frac{1-p}{p}\right)^s\cdot p\beta}{1+(1-p)\left(\frac{p}{1-p}\right)^s+
p\left(\frac{1-p}{p}\right)^s}+T=0
\end{equation}
or equivalently, defining $z=e^{\beta s}$ as the unknown, we get:
$$\frac{-(1-p)/z+pz}{1+(1-p)/z+pz}=-\frac{T}{\beta},$$
which is a quadratic equation whose relevant (positive) solution is:
$$z=z_0\eqd\frac{\sqrt{T^2+4p(1-p)(\beta^2-T^2)}-T}{2p(T+\beta)}$$
provided\footnote{Note that if $T > \beta$, the decoder will always erase (even for $R=0$)
since for $p < 1/2$, we have 
$P(\by|\bx_m)/[\sum_{m'\ne m}P(\by|\bx_{m'})] \leq (1-p)^n/p^n=e^{\beta n}
< e^{nT}$.}
that $T <\beta$, and so the derivative vanishes at
$$s_{\mbox{opt}}=s_2(p,T)\eqd
=\frac{1}{\beta}\ln\left[\frac{\sqrt{T^2+4p(1-p)(\beta^2-T^2)}-T}{2p(T+\beta)}\right].$$
It is not difficult to verify that $s_{\mbox{opt}}$ never exceeds unity.
Also, $s_{\mbox{opt}}$ is always positive 
($z_0\ge 1$) since the condition $F'(s_R) > 0$, which
is equivalent to the condition $T <\beta(p_{s_R}-p_{1-s_R})$, implies $T < \beta/2$,
which in turn is the condition for $s_{\mbox{opt}} > 0$. Note that for $T=0$, 
we obtain $s_2(p,0)=1/2$, in agreement with the Bhattacharyya bound.

In summary, the behavior of the solution can be described as follows:
As $R$ increases from $0$ to $C=\ln 2-h(p)$,
$s_R$ increases correspondingly from $0$ to $1$, and so, the expression 
$\beta(p_{s_R}-p_{1-s_R})$ 
(which is positive as long as $R < \ln 2-h(p_{1/2})$)
decreases. As long as this expression
is still larger than $T$, we have $F'(s_R) > 0$ and 
the relevant expression of $E_1^*(R,T,s)$ is $F(s)$, which is
maximized at $s=s_2(p,T)$ independently of $R$. At this range, the slope of $E_1^*(R,T)$,
as a function of $R$, is $-1$.
As $R$ continues to increase, we cross the point where $F'(s_R)=0$ 
(a point which can be thought of as an analogue to the critical 
rate of ordinary decoding) and enter into the region
where $F'(s_R) < 0$, for which $E_1^*(R,T)=G(s_1(p,R,T))$.

\subsection{More General DMC's and Random Coding Distributions}
\label{general}

Assume now a general DMC $\{P(y|x),~x\in\calX,~y\in\calY\}$ and a general i.i.d.\
random coding distribution $P(\bx)=\prod_{i=1}^n P(x_i)$ that satisfy
the condition of Theorem 1. As for
the second factor of the summand of (\ref{startingpoint}), we have the following:
\begin{eqnarray}
\bE\left\{\left(\sum_{m'\ne m} P(\by|\bX_{m'})\right)^s\right\}&=&
\bE\left\{\left(\sum_{Q_{x|y}}N_{\by}(Q_{x|y})
\cdot\exp\{n\bE_Q\ln P(Y|X)\}\right)^s\right\}\nonumber\\
&\exe&\sum_{Q_{x|y}}\bE\{N_{\by}^s(Q_{x|y})\}\cdot\exp\{ns\bE_Q\ln P(Y|X)\},
\end{eqnarray}
where $N_{\by}(Q_{x|y})$ is the number of incorrect codewords whose conditional
empirical distribution with $\by$ is $Q_{x|y}$ and
$\bE_Q$ is the expectation operator associated with $\hat{P}_{\by}\times Q_{x|y}$.
Define
$$\calG_R=\{Q_{x|y}:~R+H_Q(X|Y)+\bE_Q\ln P(x) \geq 0\},$$
where $H_Q(X|Y)$ is the conditional entropy induced by $\hat{P}_{\by}\times Q_{x|y}$.
Analogously to the case of the BSC (see also Appendix), we have:
\begin{equation}
\bE\{N_{\by}^s(Q_{x|y})\}\exe\left\{\begin{array}{ll}
\exp\{ns[R+H_Q(X|Y)+\bE_Q\ln P(x)]\} & Q_{x|y}\in\calG_R\\
\exp\{n[R+H_Q(X|Y)+\bE_Q\ln P(x)]\} & Q_{x|y}\in\calG_R^c
\end{array}\right.
\end{equation}
Thus,
\begin{eqnarray}
\bE\left\{\left(\sum_{m'\ne m} P(\by|\bX_{m'})\right)^s\right\}&\exe&
\sum_{Q_{x|y}\in\calG_R}\exp\{ns[R+H_Q(X|Y)+\bE_Q\ln P(x)]\}\times\nonumber\\
& &\exp\{ns\bE_Q\ln P(Y|X)\}+\nonumber\\
&&\sum_{Q_{x|y}\in\calG_R^c}\exp\{n[R+H_Q(X|Y)+\bE_Q\ln P(x)]\}\times\nonumber\\
& &\exp\{ns\bE_Q\ln P(Y|X)\}
\nonumber\\
&\eqd&A+B.
\end{eqnarray}
As for $A$, we obtain:
\begin{equation}
A\exe\exp\{ns[R+\max_{Q_{x|y}\in\calG_R}(H_Q(X|Y)+\bE_Q\ln[P(X)P(Y|X)])]\}
\end{equation}
Note that without the constraint $Q_{x|y}\in\calG_R$, the maximum 
of $(H_Q(X|Y)+\bE_Q\ln[P(X)P(Y|X)])$ is attained at
$$Q_{x|y}(x|y)=P_{x|y}(x|y)\eqd\frac{P(x)P(y|x)}{\sum_{x\in\calX}P(x')P(y|x')}.$$
But since $R < I(X;Y)$, then $P_{x|y}$ is in $\calG_R^c$. We argue then that the optimum
$Q_{x|y}$ in $\calG_R$ is on the boundary of $\calG_R$, i.e., it satisfies
$R+H_Q(X|Y)+\bE_Q\ln P(X)=0$. To see why this is true, consider the following
argument: Let $Q_{x|y}^0$ be any internal point in $\calG_R$ and
consider the conditional pmf $Q^t=(1-t)Q_{x|y}^0+tP_{x|y}$, $t\in[0,1]$.
Define $f(t)=H_{Q^t}(X|Y)+\bE_{Q^t}\ln[P(X)P(Y|X)]$. Obviously, $f$ is concave and
$f(0)\le f(1)$. Now, since $Q^0\in\calG_R$ and $Q^1=P_{x|y}\in\calG_R^c$, then
by the continuity of the function $R+H_{Q^t}(X|Y)+\bE_{Q^t}\ln P(X)$, there must be
some $t=t_0$ for which $Q^{t_0}$ is on the boundary of $\calG_R$. By the concavity of $f$,
$f(t_0)\ge(1-t_0)f(0)+t_0f(1)\ge f(0)$. Thus, any internal point of $\calG_R$ can be
improved by a point on the boundary between $\calG_R$ and $\calG_R^c$.
Therefore, we have
\begin{eqnarray}
& &\max_{Q_{x|y}\in\calG_R}(H_Q(X|Y)+\bE_Q\ln[P(X)P(Y|X)])]\nonumber\\
&=&\max_{\{Q_{x|y}:~H_Q(X|Y)+\bE_Q\ln P(X)=-R\}}[H_Q(X|Y)+\bE_Q\ln P(X)+\bE_Q\ln P(Y|X)]\nonumber\\
&=&\max_{\{Q_{x|y}:~H_Q(X|Y)+\bE_Q\ln P(X)=-R\}}[-R+\bE_Q\ln P(Y|X)]\nonumber\\
&=&-R+\max_{\{Q_{x|y}:~H_Q(X|Y)+\bE_Q\ln P(X)=-R\}}\bE_Q\ln P(Y|X)\nonumber\\
&=&-R+\max_{Q_{x|y}\in\calG_R}\bE_Q\ln P(Y|X)
\end{eqnarray}
which means that $A\exe e^{-ns\Delta(R)}$, where
$$\Delta(R)=\min_{Q_{x|y}\in\calG_R}\bE_Q\ln[1/P(Y|X)].$$
The achiever of $\Delta(R)$ is of the form
$$Q(x|y)=\frac{P(x)P^{s_R}(y|x)}{\sum_{x'\in\calX}P(x')P^{s_R}(y|x')},$$
where $s_R$ is such that $H_Q(X|Y)+\bE_Q\ln P(X)=-R$,
or equivalently, $s_R$ is the solution 
\footnote{Observe that for $s=0$, $H_Q(X|Y)+\bE_Q\ln P(X)=0$ and for $s=1$,
$H_Q(X|Y)+\bE_Q\ln P(X)=-I(X;Y) < -R$. Thus for $R < I(X;Y)$, $s_R\in[0,1)$.}
to the equation $s\gamma'(s)-\gamma(s)=R$.
In other words,
$$\Delta(R)=\frac{\sum_{x\in\calX}P(x)P^{s_R}(y|x)\ln[1/P(y|x)]}{
\sum_{x\in\calX}P(x)P^{s_R}(y|x)}=\gamma'(s_R).$$

Considering next the expression of $B$, we have:
$$B\exe\exp\{n[R+\max_{Q_{x|y}\in\calG_R^c}(H_Q(X|Y)+\bE_Q\ln P(X)+s\bE_Q\ln P(Y|X))]\}.$$
The unconstrained maximizer of $(H_Q(X|Y)+\bE_Q\ln P(X)+s\bE_Q\ln P(Y|X))$ is
$$Q_{x|y}^{(s)}(x|y)=\frac{P(x)P^s(y|x)}{\sum_{x'\in\calX}P(x')P^s(y|x')}.$$
Now, there are two cases, depending on the value of $s$: If $s$
is such that $Q_{x|y}^{(s)}\in\calG_R^c$, or equivalently, $s > s_R$,
then $B\exe e^{-n[\gamma(s)-R]}$. 
If $Q_{x|y}^{(s)}\in\calG_R$, namely, $s\leq s_R$, then once again, the optimum is attained at the
boundary between $\calG_R$ and $\calG_R^c$, and then $B\exe e^{-ns\gamma'(s_R)}$.
In summary, $B\exe e^{-n\Lambda(R,s)}$, where
$$\Lambda(R,s)=\left\{\begin{array}{ll}
\gamma(s)-R & s > s_R\\
s\gamma'(s_R) & s \leq s_R
\end{array}\right. $$
The dominant term between $A$ and $B$ is obviously always $B$ because it is either
of the same exponential order of $A$, in the case $s\leq s_R$, or has a slower exponential
decay, when $s > s_R$, as then the global (unconstrained) maximum 
of $[H_Q(X|Y)+\bE_Q\ln P(X)+s\bE_Q\ln P(Y|X)]$ is achieved.
Thus, putting it all together, we get:
\begin{eqnarray}
\overline{\mbox{Pr}}\{\calE_1\}&\lexe& e^{nsT}\cdot|\calY|^n\cdot e^{-n\gamma(1-s)}\cdot
e^{-n\Lambda(R,s)}\nonumber\\
&=& e^{-nE_1^*(R,T,s)}
\end{eqnarray}
and the optimum $s\ge 0$ gives $E_1^*(R,T)$.

\clearpage
\section*{Appendix}
\renewcommand{\theequation}{A.\arabic{equation}}
    \setcounter{equation}{0}

We begin with a simple large deviations bound regarding
the distance enumerator. In fact, this bound (in a slightly different form)
was given already in \cite{p117}, but we present here too for the sake of
completeness.
For $a,b\in[0,1]$, consider the binary divergence
\begin{eqnarray}
D(a\|b)&\eqd&a\ln \frac{a}{b}+(1-a)\ln\frac{1-a}{1-b}\nonumber\\
&=&a\ln \frac{a}{b}+(1-a)\ln\left[1+\frac{b-a}{1-b}\right].
\end{eqnarray}
To derive a lower bound to $D(a\|b)$, let us use the inequality
\begin{equation}
\label{lnineq}
\ln(1+x)=-\ln\frac{1}{1+x}=-\ln\left(1-\frac{x}{1+x}\right)\ge \frac{x}{1+x},
\end{equation}
and then
\begin{eqnarray}
D(a\|b)&\ge&a\ln \frac{a}{b}+(1-a)\cdot\frac{(b-a)/(1-b)}
{1+(b-a)/(1-b)}\nonumber\\
&=&a\ln \frac{a}{b}+b-a\nonumber\\
&>&a\left(\ln\frac{a}{b}-1\right).
\end{eqnarray}
Consider first the binary case (the extension to the general case is straightforward
as will be explained below).
For every given $\by$, $N_{\by}(d)$ is the sum of the $e^{nR}-1$
independent binary random variables, $\{1\{d(\bX_{m'},\by)=d\}\}_{m'\ne m}$,
where the probability that $d(\bX_{m'},\by)=n\delta$ is
exponentially $b=e^{-n[\ln 2- h(\delta)]}$. 
The event $N_{\by}(n\delta)\ge e^{nA}$, for $A\in[0,R)$,
means that the relative frequency of the event $1\{d(\bX_{m'},\by)=n\delta\}$
is at least $a=e^{-n(R-A)}$. Thus, by the Chernoff bound:
\begin{eqnarray}
\mbox{Pr}\{N_{\by}(n\delta)\ge e^{nA}\}&\lexe&
\exp\left\{-(e^{nR}-1)D(e^{-n(R-A)}\|e^{-n[\ln 2-
h(\delta)]})\right\}\nonumber\\
&\lexe&
\exp\left\{-e^{nR}\cdot e^{-n(R-A)}(n[(\ln 2-R-
h(\delta)+A]-1)\right\}\nonumber\\
&\le&
\exp\left\{-e^{nA}(n[\ln 2-R-h(\delta)+A]-1)\right\}.
\end{eqnarray}
Therefore, for $\delta\in\calG_R^c$, we have:
\begin{eqnarray}
\bE\{N_{\by}^s(n\delta)\}&\le&e^{n\epsilon s}\cdot\mbox{Pr}\{1\le N_{\by}(n\delta)\le e^{n\epsilon}\}+
e^{nRs}\cdot \mbox{Pr}\{ N_{\by}(n\delta)\ge e^{n\epsilon}\}\nonumber\\
&\le&e^{n\epsilon s}\cdot\mbox{Pr}\{N_{\by}(n\delta)\ge 1\}+
e^{nRs}\cdot \mbox{Pr}\{ N_{\by}(n\delta)\ge e^{n\epsilon}\}\nonumber\\
&\le&e^{n\epsilon s}\cdot\bE\{N_{\by}(n\delta)\}+
e^{nRs}\cdot e^{-(n\epsilon-1)e^{n\epsilon}}\nonumber\\
&\le&e^{n\epsilon s}\cdot e^{n[R+h(\delta)-\ln 2]}+
e^{nRs}\cdot e^{-(n\epsilon-1)e^{n\epsilon}}.
\end{eqnarray}
One can let $\epsilon$ vanish with $n$ sufficiently slowly that
the second term is still superexponentially small, e.g., $\epsilon=1/\sqrt{n}$.
Thus, for $\delta\in\calG_R^c$, $\bE\{N_{\by}^s(n\delta)\}$ is exponentially bounded
by $e^{n[R+h(\delta)-\ln 2]}$ independently of $s$. For $\delta\in\calG_R$, we have:
\begin{eqnarray}
\bE\{N_{\by}^s(n\delta)\}&\le&e^{ns[R+h(\delta)-\ln 2+\epsilon]}\cdot
\mbox{Pr}\{N_{\by}(n\delta)\le e^{n[R+h(\delta)-\ln 2+\epsilon]}\}+\nonumber\\
& &e^{nRs}\cdot
\mbox{Pr}\{N_{\by}(n\delta)\ge e^{n[R+h(\delta)-\ln 2+\epsilon]}\}\nonumber\\
&\le&e^{ns[R+h(\delta)-\ln 2+\epsilon]}
+e^{nRs}\cdot e^{-(n\epsilon-1)e^{n\epsilon}}
\end{eqnarray}
where again, the second term is exponentially negligible.

To see that both bounds are exponentially tight, consider
the following lower bounds. For $\delta\in\calG_R^c$,
\begin{eqnarray}
\bE\{N_{\by}^s(n\delta)\}&\ge&1^s\cdot\mbox{Pr}\{N_{\by}(n\delta)=1\}\nonumber\\
&=&e^{nR}\cdot\mbox{Pr}\{d_H(\bX,\by)=n\delta\}\cdot
\left[1-\mbox{Pr}\{d_H(\bX,\by)=n\delta\}\right]^{e^{nR}-1}\nonumber\\
&\exe&e^{nR}e^{-n[\ln 2-h(\delta)]}\cdot
\left[1-e^{-n[\ln 2-h(\delta)]}\right]^{e^{nR}}\nonumber\\
&=&e^{n[R+h(\delta)-\ln 2]}\cdot\exp\{e^{nR}\ln[1-e^{-n[\ln 2-h(\delta)]}]\}.
\end{eqnarray}
Using again the inequality in (\ref{lnineq}),
the second factor is lower bounded by
$$\exp\{-e^{nR}e^{-n[\ln 2-h(\delta)]}/(1-e^{-n[\ln 2-h(\delta)]})\}
=\exp\{-e^{-n[\ln 2-R-h(\delta)]}/(1-e^{-n[\ln 2-h(\delta)]})\}$$
which clearly tends to unity as $\ln 2-R-h(\delta) > 0$ for $\delta\in\calG_R^c$.
Thus, $\bE\{N_{\by}^s(n\delta)\}$ is exponentially lower bounded by $e^{n[R+h(\delta)-\ln 2]}$.
For $\delta\in\calG_R$, and an arbitrarily small $\epsilon > 0$, we have:
\begin{eqnarray}
\bE\{N_{\by}^s(n\delta)\}&\ge& e^{ns[R+h(\delta)-\ln 2-\epsilon]}\cdot
\mbox{Pr}\{N_{\by}(n\delta)\ge e^{n[R+h(\delta)-\ln 2-\epsilon]}\}\nonumber\\
&=& e^{ns[R+h(\delta)-\ln 2-\epsilon]}\cdot\left(1-
\mbox{Pr}\{N_{\by}(n\delta)< e^{n[R+h(\delta)-\ln 2-\epsilon]}\}\right)
\end{eqnarray}
where $\mbox{Pr}\{N_{\by}(n\delta)< e^{n[R+h(\delta)-\ln 2-\epsilon]}\}$ is again
upper bounded, for an internal point in $\calG_R$,
by a double exponentially small quantity as above. 
For $\delta$ near the boundary of $\calG_R$, namely, when $R+h(\delta)-\ln 2\approx 0$,
we can lower bound $\bE\{N_{\by}^s(n\delta)\}$ by slightly reducing $R$ to $R'=R-\epsilon$ 
(where $\epsilon > 0$ is very small). This will make
$\delta$ an internal point of $\calG_{R'}^c$ for which the previous bound applies, 
and this bound
is of the exponential order of $e^{n[R'+h(\delta)-\ln 2]}$.
Since $R'+h(\delta)-\ln 2$ is still very close to zero, then $e^{n[R'+h(\delta)-\ln 2]}$
is of the same exponential order as $e^{ns[R+h(\delta)-\ln 2]}$
since both are about $e^{0\cdot n}$.

The above proof extends straightforwardly from the binary case to the more general case.
The only difference is that in the general case,
for a given $\by$, the probability that a random codeword,
drawn under $\{P(x)\}$, would have a given conditional empirical distribution $Q_{x|y}$
with $\by$, is exponentially $e^{n[H_Q(X|Y)+\bE_Q\ln P(X)]}$. Thus, $h(\delta)-\ln 2$
of the binary case has to be replaced by $H_Q(X|Y)+\bE_Q\ln P(X)$ in all places.

\end{document}